\documentclass[twocolumn,showpacs,amssymb]{revtex4}
\usepackage{graphicx}

\newcommand{\eV}{\rm{\, eV }}

\newcommand{\MeV}{\rm{\, MeV }}
\newcommand{\GeV}{\rm{\, GeV }}

\newcommand{\beq}{\begin{equation}}
\newcommand{\eeq}{\end{equation}}
\newcommand{\ba}{\begin{array}}
\newcommand{\ea}{\end{array}}

\newcommand{\ee}{\epsilon_{e,0}}

\def \etal{{\it et al.~}}
\def\be{\begin{equation}}
\def\ee{\end{equation}}
 \def\lsim{\raisebox{-0.3ex}{\mbox{$\stackrel{<}{_\sim} \,$}}}
 \def\gsim{\raisebox{-0.3ex}{\mbox{$\stackrel{>}{_\sim} \,$}}}

\begin{document}
\title{Radio Quiet AGNs as Possible Sources of Ultra-high Energy
  Cosmic Rays}


\author {Asaf Pe'er$^{1,2}$, Kohta Murase$^3$ and Peter M\'esz\'aros$^4$}
\affiliation{
$^1$Space Telescope Science Institute, 
Baltimore, MD 21218, USA\\
$^2$Riccardo Giacconi Fellow\\
$^3$Yukawa Institute for Theoretical Physics, Kyoto University, Kyoto,
606-8502, Japan\\
$^4$Department of Astronomy \& Astrophysics; Department of Physics; Center for
Particle Astrophysics; Pennsylvania State University, University Park,
PA 16802, USA
}


\begin{abstract}
  Active galactic nuclei (AGNs) have been one of the most widely
  discussed sources of ultrahigh-energy cosmic rays (UHECRs).  The
  recent results of Pierre Auger observatory (PAO) have indicated a
  possible composition change of UHECRs above $\sim10^{18.5}$~eV
  towards heavy nuclei.  We show here that if indeed UHECRs are
  largely heavy nuclei, then nearby radio quiet AGNs can also be
  viable sources of UHECRs.  We derive constraints on the acceleration
  sites which enable acceleration of UHECRs to $10^{20}$~eV without
  suffering losses.  We show that the acceleration of UHECRs and the
  survival of energetic heavy nuclei are possible in the parsec scale
  weak jets that are typically observed in these objects, the main
  energy loss channel being photodisintegration. On this scale, energy
  dissipation by shock waves resulting from interactions inside a jet
  or of the jet with surrounding material are expected, which may
  accelerate the particles up to very high energies.  We discuss the
  possible contribution of radio-quiet AGNs to the observed UHECR
  flux, and show that the required energy production rate in UHECRs by
  a single object could be as low as $\approx 3 \times 10^{39} \rm{\,
    erg \, s^{-1}}$, which is less than a percent of the bolometric
  luminosity, and thus energetically consistent. We discuss
  consequences of this model, the main one being the difficulty in
  detecting energetic secondaries ($\gamma$-rays and neutrinos) from
  the same sources.

\end{abstract}


\maketitle


\section{Introduction}
\label{sec:intro}

The origin of ultrahigh-energy cosmic rays (UHECRs), cosmic rays with
energies above $\gtrsim 10^{18.5}$~eV, is still a mystery. It is
commonly believed that the sources of UHECR's are extragalactic, since
at these energies the cosmic rays cannot be confined by the magnetic
field of our galaxy (for recent reviews, see Refs. \cite{Der07,
  BEH09}).  Due to energy losses by photo-meson production with the
cosmic microwave background (CMB), the sources of UHECR's are further
limited to a distance $\lsim 100$~Mpc from the observer (this limit is
known as the `Greisen-Zatsepin-Kuzmin (GZK) cutoff';
Refs. \cite{Greisen66, ZK66}).  While early observations \cite{Tak98}
suggested a violation of the GZK cutoff at high energies, recent, high
statistics observations by the High-Resolution Fly's eye (HiRes)
experiment and the Pierre Auger observatory (PAO) finds a downturn in
the spectrum consistent with the GZK predictions \cite{Abbasi08,
  Abraham08b}.

Extracting information from observations is, however, not easy. For
example, observations of the spectrum itself did not lead, so far, to
a firm conclusion about the origin of the steepening at high energies
(i.e., whether it originates from the GZK cutoff or not), and
therefore other possibilities are still allowed.

One of the consequences of this difficulty is the lack of a clear
theoretical picture of the extragalactic UHECR sources.  Possible
candidates of UHECR sources are limited among the known astrophysical
objects, because of the difficulty in satisfying the two nearly
contradictory requirements: A strong magnetic field in the source is
needed to confine the accelerated cosmic rays, while the magnetic and
photon fields cannot be too strong in order to avoid too much
synchrotron radiation and photo-meson energy losses \cite{Hillas84}.
Several possible sources of UHECRs that fulfill the constraints
mentioned above are often discussed in the literature. The most widely
discussed candidates are gamma ray bursts (GRB )\cite{MU95, W95,
  Vietri95, Der02, W04, MINN08}, low luminosity GRBs and hypernovae
\cite{MINN06, WRMD07, MINN08}, and jets in radio-loud (RL) AGNs
\cite{BS87, RB93, NMA95, Aha02, BGG06, Berezhko08, DRFA08}.  UHECR
acceleration in the vicinity of black holes has also been considered
for AGNs, including radio-quiet (RQ) AGNs (see, e.g.,
Refs. \cite{PS92,BG99}).  Additional suggestions include magnetars
\cite{Arons03,MMZ09} and clusters of galaxies \cite{KRJ96,ISMA07}.

An important step towards understanding the origin of UHECRs was taken
recently with the discovery of a correlation between the arrival
direction of the highest energy cosmic rays and nearby galaxies in the
\cite{VCV06} (VCV) catalog of active galactic nuclei (AGNs) [12th
edition] \cite{Abraham07, Abraham08a}. These results, however, should
be taken with great care. It was noted already by Refs. \cite{Abraham07,
  Abraham08a} that the observed correlation alone is insufficient to
conclude that the sources of UHECRs are AGNs, because AGNs themselves
are clustered within galaxies, and thus may only be the tracers of the
true sources.  Indeed, a correlation was found between the arrival
directions of UHECRs and galaxies \cite{KS08,Tak08,Ghis08}, which
suggests a correlation with the large scale structure (LSS) in the
nearby universe.

The emerging picture gets further complicated by the recent results of
HiRes experiment which do not confirm the correlation found by PAO
\cite{Abbasi08b}.  Moreover, the latest PAO results \cite{Abraham09}
suggest that the correlation degree is reduced compared to that
obtained by earlier measurements.

An additional clue on the nature of UHECRs sources may come from
measurements of their chemical composition. However, deducing the
chemical composition of UHECRs from current observations is also
difficult due to our poor knowledge on hadronic interactions. PAO
results indicate a heavy composition at the highest energies
\cite{Unger07, Belido09}, while HiRes results suggest a proton
composition (see Ref. \cite{BEH09}, and references therein). The
results of these two experiments are therefore clearly in
contradiction. Moreover, the heavy element composition suggested by
PAO results may be in contradiction with the anisotropy result of the
same experiment (see below).

The chemical composition of UHECRs may be crucial in determining the
nature of their source. If UHECRs are protons, the theoretical
restrictions on the physical conditions inside potential sources are
rather extreme, so that if indeed AGNs are the sources of UHECRs, then
only powerful AGNs such as RL AGNs can be viable sources.  In
addition, the known values of the galactic magnetic field, as well as
observational upper limits on the intergalactic magnetic fields
(IGMFs) in voids \cite{Kro94} and theoretical investigations about
IGMFs in the structured region \cite{RKCD08} allow one to expect a
correlation between the arrival directions of UHECRs and AGNs
\cite{DKRC08}, which is tentatively suggested by PAO.  On the other
hand, a drawback of the idea that AGNs are sources of UHECRs is the
fact that a strong correlation with RL AGNs is not found, except for
very few cases, such as Cen A or Cen B \cite{MSPC09}. In addition, a
correlation with the most powerful AGNs, FR II galaxies, is currently
not confirmed. Moreover, the number density of FR II galaxies seems
too small to explain the observed small scale anisotropy \cite{TS09}.
Furthermore, the observed bolometric luminosity of AGNs which are
correlated with the arrival direction of UHECRs seems too small to
satisfy the minimum conditions for UHECRs acceleration in continuous
jets \cite{ZFG09}. These results indicate that only very few nearby
AGNs are sufficiently powerful to accelerate particles to the observed
ultrahigh energies.

In order to overcome this problem, it was suggested by Ref. \cite{FG09}
that UHECR production may occur transiently during giant AGN flares
caused by tidal disruptions.  Although this may be a valid scenario,
several restrictions on the duration and rate of such flares can be
derived \cite{Sig09,MT09}.  Since the actual activity of AGNs is not
known, concrete conclusions cannot be drawn at this stage (see
also Ref. \cite{WL08}).

An alternative picture emerges if UHECRs are composed of heavier
elements, such as iron nuclei. The ambiguity in the interpretation of
the data led \cite{Allard05, Allard08} to study a mixed composition
scenario above $\sim 10^{18.5}$~eV. A similar model (albeit with
somewhat lower maximum energy of UHECR) was studied recently by
\cite{ABG09}.  In these models, significant correlation between the
arrival directions of UHECRs and galaxies is not expected, because of
the strong deflection of heavy nuclei in galactic magnetic field. In
addition, protons dominate the cosmic rays spectra only up to energies
$\sim {10}^{18.5}$~eV, which implies that the requirement on the
physical conditions at the acceleration site of potential sources is
significantly loosened.

As we show here, a heavy element composition of UHECRs opens the
possibility for RQ AGNs, which compose the majority of the AGN
population within $\sim 100$~Mpc, to be the sources of UHECRs.  As
opposed to RL AGNs, in which evidence exist for strong jets on
$\gtrsim$~kpc scale, in RQ AGNs there are no evidence of such strong
jets \footnote{Note though that \cite{PC09} showed recently that radio
  emission from jets may be suppressed if the emitting particles cool
  rapidly enough close to the base of the jet, so that synchrotron
  emission from outer parts of the jet is self absorbed. Although this
  work was done in the context of jets in microquasars, a similar idea
  can be easily implemented in the case of RQ AGNs as well. However, a
  general discussion on the role of synchrotron-self absorption on the
  distinguishing between RL and RQ AGN's is outside the scope of this
  paper, and is left for future work.}. Nonetheless, in recent years
there has been an accumulation of evidence for weak, parsec-scale jets
in these objects (Refs. \cite{MWPG03,GBO04,UWTGM05,Gal+06,
  MARK07,Ho08}, and references therein).  Internal interactions of
blobs inside a jet, or of the jets with their environment, produce
shock waves, which can accelerate particles.  Alternatively, cosmic
rays may be accelerated by the dissipation of magnetic energy inside
these jets (see, e.g., Ref. \cite{LO07}). Moreover, as was pointed out
by Ref. \cite{AM04}, particle acceleration could take place in shocks
within the corona, which could be caused by abortive jets.

In this work we thus discuss weak jets in RQ AGNs as possible sources
of UHECRs, under the assumption that at very high energies the
composition of UHECRs is dominated by heavy (possibly, but not
necessarily, iron) nuclei. As we will show below, RQ AGNs can fulfill
the requirements needed from sources of UHECRs under this
assumption. We assume here that particles acceleration takes place
inside the jets, so that this work is different from previous works
that considered UHECR production in the vicinity of black holes that
do not have jets (e.g., Ref. \cite{BG99}).

This paper is organized as follows. In \S\ref{sec:2} we present the
basic constraints on the physical conditions at the acceleration sites
of heavy nuclei. In particular, in \S\ref{sec:photodisintegration} we
derive the constraints from photodisintegration and photomeson
production. We show that these conditions can be fulfilled easily.
In \S\ref{sec:3} we discuss the efficiency of particle acceleration
required to explain the observed flux of UHECRs on earth, under the
assumption that RQ AGNs are the only sources of UHECRs.  We further
derive constraints on the value of the magnetic field at the
acceleration site, and show that a lower limit of few percents of
equipartition value is required. We summarize and conclude in
\S\ref{sec:summary}.

\section{Constraints on the Emission Site of UHECRs}
\label{sec:2}

The acceleration sites of UHECRs are required to fulfill two basic
conditions (see, e.g., Refs. \cite{Hillas84, NMA95, W95}).  The first
condition is that the accelerated particles should be confined to the
acceleration region (this is also known as the ``Hillas condition'',
see Ref. \cite{Hillas84}). Assuming that the acceleration results from
electromagnetic processes within an expanding plasma, this condition
is equivalent to the requirement that the acceleration time is shorter
than the dynamical time scale.  Considering jetted plasma, which moves
relativistically with velocity $\beta c$, or Lorentz factor $\Gamma =
(1-\beta^2)^{-1/2}$, the acceleration time scale in the comoving frame
can be written as $t_{\rm acc} \simeq \eta E^{ob} /(\Gamma Z q B c)$
\footnote{Note that we adopt here the common assumption that the
  particle velocity in the shock frame is comparable to the shock
  velocity.}. Here, $E^{ob} = \Gamma E'$ is the observed energy of the
particle \footnote{Although we may not expect relativistic motion in
  the weak jets of radio quite AGNs, so that $\Gamma \gtrsim 1$, we
  introduce this quantity here for completeness.}, $E'$ is its energy
in the comoving frame, $B$ is the magnetic field, and $Z q$ is the
nucleon charge.  The factor $\eta \geq 1$ is a dimensionless factor,
whose exact value is determined by the (yet uncertain) details of the
acceleration mechanism. For example, in the non-relativistic diffusive
shock acceleration mechanism, this factor corresponds to $\eta =
(20/3) \beta^{-2}$ in the Bohm limit for parallel shocks (e.g.,
Refs. \cite{Drury83, BE87}).  For relativistic shocks, $\eta \sim$ a
few can be expected but larger values are also possible. The condition
$t_{\rm acc} < t_{\rm dyn} = r/\Gamma \beta c$, gives
\beq
B \geq {\eta E^{ob} \beta \over Z q r} = { 3.3 \times 10^{17} \eta
  \beta \over r} \,
Z^{-1} \; {E_{20}^{ob}} \; \rm{G},
\label{eq:1}
\eeq
where $r$ is the radial distance from the source at which particle
acceleration takes place, measured in cm. Here and below, we use the
convention $Q = 10^X Q_X$ in cgs units. 

Writing the energy density in the magnetic field as a fraction
$\epsilon_B$ of the photon energy density at the acceleration site
$u_B \equiv B^2/8 \pi = \epsilon_B u$, where $u = L/(4\pi r^2 \Gamma^2
\beta c)$ is the photon energy density (representative of the electron
energy density) and $L$ is the photon luminosity, equation \ref{eq:1}
can be written in the form (Ref. \cite{FG09})
\beq
\epsilon_B L \geq 1.7 \times 10^{45} \, \eta^2 \, Z^{-2} \, \Gamma^2
\, \beta^{3} \; {E_{20}^{ob}}^2 \; \rm{erg \, s^{-1}}.
\label{eq:2}
\eeq
Equations \ref{eq:1}, \ref{eq:2} give basic conditions required from a
source capable of accelerating particles to the maximum observed UHECR
energy, $\sim 10^{20}$~eV. However, as the details of the
acceleration process are uncertain, the actual restriction may depend
on additional phenomena, such as the details of particle escape from
the acceleration region. The details of particle escape are unknown,
and the maximum energy may be escape-limited if the escape time
scale is shorter than the dynamical time scale. In this paper, we simply
assume that acceleration is efficient, such that $t_{\rm esc} \gtrsim
t_{\rm dyn}$, which can be expected when the Bohm limit is achieved
over the size of $r/\Gamma$.
 
For protons (Z=1), equation \ref{eq:2} necessitates a magnetic
luminosity which is 1-2 orders of magnitude above the bolometric
photon luminosity observed in nearby AGNs associated with UHECR's in
the VCV catalog, $L_{bol}^{ob} \approx 10^{44} {\rm erg \, s^{-1}}$
\cite{ZFG09}. We therefore expect that for proton composition of
UHECR, acceleration to the highest energies is possible only in
powerful RL AGNs (for which $L \gtrsim {10}^{44}~{\rm erg} \, {\rm
  s}^{-1}$). However, for heavier nuclei, equation \ref{eq:2} implies
a much weaker constraint: e.g., for Carbon nuclei ($Z=6$) one obtains
$\epsilon_B L \geq 5 \times 10^{43} \, \eta^2 \, \Gamma^2 \, \beta^{3}
\; {E_{20}^{ob}}^2 \; \rm{erg \, s^{-1}}$.  For iron nuclei, $Z=26$
and the constraint on the luminosity further eases to $\epsilon_B L
\geq 2.5 \times 10^{42} \, \eta^2 \, \Gamma^2 \, \beta^{3} \;
{E_{20}^{ob}}^2 \; \rm{erg \, s^{-1}}$, well within the limits of many
nearby AGNs' observed luminosity (except for low-luminosity AGNs, with
$L \lsim 10^{42} \; \rm{erg \, s^{-1}}$) \footnote{ Additional
  possible solutions to the problem of high luminosity presented in
  equation \ref{eq:2} is obtained either (1) by assuming that the
  energy density in the magnetic field {\it at the acceleration site}
  is larger by 1-2 orders of magnitude than the energy density in the
  radiating electrons.  In such a scenario, acceleration of protons to
  the maximum observed energy of $10^{20}$~eV is also acceptable.(2) A
  second possibility is large luminosity flaring activity, as might
  have seen by comparing {\it FERMI} to earlier {\it EGRET}
  observations \cite{Abdo+09}.}.

The second condition is that the acceleration time is shorter than all
the relevant energy loss time scales such as synchrotron loss time.
For the synchrotron loss time, we have $t_{\rm cool, syn} = (6 \pi
m_p^4 c^3 \Gamma A^4)/(\sigma_T m_e^2 B^2 E^{ob} Z^4)$ (see,
e.g., Ref. \cite{WRM08}).  Here, $\sigma_T$ is Thomson's cross section,
$m_p$ and $m_e$ are the proton and electron masses and $A m_p$ is the
mass of the nucleon (for iron nuclei, $A=56$ and $Z=26$). The
requirement $t_{\rm acc} < t_{\rm cool, syn}$ results in
\beq
E^{ob} \leq 2 \times 10^{20} \; \Gamma\, \eta^{-1/2}\,  B^{-1/2}\, A^2
\, Z^{-3/2} \eV.  
\label{eq:3}
\eeq
For iron nuclei, equation \ref{eq:3} gives $E^{ob} \leq 5 \times
10^{21} \; \Gamma \, \eta^{-1/2} \, (B/1 {\rm G})^{-1/2} \eV$, which
implies that in order to enable acceleration of cosmic rays to the
highest observed energies, $\sim 10^{20}$~eV, the strength of the
magnetic field at the acceleration site should not exceed $B \,
\lsim$~few - few tens G (as long as $\Gamma \sim 1$, as is
expected in RQ AGNs. See Ref. \cite{Gal+06}).

The constraints of equations \ref{eq:1} and \ref{eq:3} can be combined
to constrain the radius of the acceleration site, 
\beq
r_{\rm acc} \equiv r \geq 9 \times 10^{16} \,  \eta^2 \, \Gamma^{-2} \, \beta
\, Z^{2} \, A^{-4} \; {E_{20}^{ob}}^3 \; {\rm cm}.
\label{eq:4}
\eeq
For iron nuclei, equation \ref{eq:4} implies $r_{acc} \geq 6\times
10^{12} \, \eta^2 \, \Gamma^{-2} \, \beta \, {E_{20}^{ob}}^3 \; {\rm
cm}$, which is comparable to the Schwarzshield radius of typical black
holes in AGNs (with characteristic mass $M_{BH} = 10^8 M_\odot$),
$r_{Sch, M_8} = 3 \times 10^{13}$~cm.

Comparison of the synchrotron cooling time and the dynamical time
shows that if the acceleration takes place at radii smaller than 
\beq
r_{\rm acc} \leq {1.8 \times 10^{18} \over B^2} \, \Gamma^2 \, \beta \, (A/Z)^4
  \, {E_{20}^{ob}}^{-1} \; {\rm cm},
\label{eq:t_cool_dyn}
\eeq
then the synchrotron cooling time for the most energetic particles is
shorter than the dynamical time, $t_{\rm cool, syn}(E_{20}^{ob}=1) <
t_{\rm dyn}$. Assuming that the particle escape time is comparable to
the dynamical time, when the condition in equation \ref{eq:t_cool_dyn}
is met, it implies that the spectrum of escaped particles is similar
to that determined by the acceleration mechanism. We further discuss
the implication of this condition in section \ref{sec:3}
below. Combining equations \ref{eq:4} and \ref{eq:t_cool_dyn}, one obtains an
upper limit on the value of the magnetic field at the acceleration site,
\beq
B \leq 4  \, \Gamma^2 \,  A^4 \, Z^{-3}  \, {E_{20}^{ob}}^{-2} \, \eta^{-1} \; {\rm G}.
\label{eq:B_max}
\eeq
A lower limit on the value of the magnetic field can only be obtained
once the acceleration radius $r$ is specified. This will be 
discussed in section \ref{sec:3} below.

In addition to synchrotron energy losses, energetic particles can in
principle lose their energy by interacting with the ambient photon
field and with other nuclei.  Interaction with the photon field can
result in energy losses by Compton scattering (which results in
negligible energy losses for UHECR's, and therefore will not be
considered here), photopair production (Bethe-Heitler process),
photo-production of mesons (mainly pions), and photodisintegration of
the nuclei. 
Therefore, the condition $t_{\rm
  acc}<t_{\rm cool}$ as given in equation \ref{eq:2} should be
modified.

\subsection{Constraints from Disintegration of Energetic Nuclei}
\label{sec:photodisintegration}

The threshold energy for photopair production, $\epsilon_{\rm th}^\pm
= 2 m_e (1+m_e/m_p) \simeq 1 \MeV$, is lower than the threshold energy
for pion production, $\epsilon_{\rm th}^\pi = m_\pi(1+m_\pi/2 m_p)
\simeq 145 \MeV$. The relative contributions of these two processes to
the energy loss of a relativistic particle is determined by the
product of the cross section and inelasticity coefficient of the
relevant process.  For energetic protons, the product of cross section
and inelasticity in pion production process is two orders of magnitude
larger than the product in pair creation process \cite{BRS90,CZS92}.
Exact calculation must take into account the target photon
spectrum. It can be shown that for a photon number spectral index
$\alpha=2$ photomeson production is somewhat more dominant, however
both time scales are comparable \cite{MPR01}.  For heavy nuclei this
ratio drops by a factor $Z^2/A$ (see Refs. \cite{BRS90, AHST08} and
references therein), which for iron nuclei still implies that
photomeson production is more important than photopair production as
an energy loss channel.

However, the main energy loss channel of energetic particles is
photodisintegration \cite{PSB76}. The threshold energy for this
process is $\epsilon_{\rm th}^A \simeq 10 \MeV$, larger than the
threshold energy for photopair production.  Here, we thus focus on
photomeson production and photodisintegration as the main energy loss
channels of energetic particles interacting with the photon field.

An energetic nucleus having Lorentz factor $\gamma_A$ (in the plasma
comoving frame) propagating through an isotropic photon background
with differential number density $n(\varepsilon_\gamma)
d\varepsilon_\gamma$ (at the energy range $\varepsilon_\gamma
... \varepsilon_\gamma + d\varepsilon_\gamma$) loses energy by
photodisintegration and photomeson production at a rate
(see, e.g., Refs. \cite{Stecker68, Stecker69})
\beq
t_{{\rm dis},\pi}^{-1} = {c \over 2 \gamma_A^2} \int_{\epsilon_{\rm th}^{A,\pi}}^\infty 
\epsilon' \sigma_{A,\pi}(\epsilon') \kappa(\epsilon') d\epsilon'
\int_{\epsilon'/2 \gamma_A}^\infty {n(\varepsilon'') \over
  \varepsilon''^2} d\varepsilon''.
\label{eq:t_dis1}
\eeq
Here, $\sigma_{A,\pi}(\epsilon')$ are the cross sections for
photodisintegration and photomeson production, and $\kappa(\epsilon')$
is the inelasticity coefficient in the photomeson production
process. In calculating survival of one specie of heavy nuclei due to
the photodisintegration process, equation \ref{eq:t_dis1} can still be
used by inserting $\kappa(\epsilon') \equiv 1$.

We estimate the differential photon number density at the acceleration
radius $r_{\rm acc}$ in the following way: as a rough approximation,
the spectrum of RQ AGNs increases as $\nu L_\nu \propto \nu^2$ at low
energies, $\nu< \nu_0^{ob} \equiv 10^{13}$~Hz, and is flat ($\nu L_\nu
\propto \nu^0$) at higher frequencies, up to the X-rays
($\nu_{\max}^{ob} \sim 10^{18}$~Hz; See Refs. \cite{Elvis94, RR05,
  Richards06}) \footnote {The actual spectrum is of course much more
  complicated than this simple estimate. The radio emission may
  originate from weak jets, infrared emission may come from AGN tori
  while UV and x-ray emission are typically radiated by the accretion
  disks and coronae. The estimation given here is therefore mainly
  valid for checking the possibility of survival of UHE nuclei, but
  may not be valid for more accurate predictions of hadronic
  emission.}.  The differential photon number density can therefore be
written as a broken power law,
\beq
n(\varepsilon) = n_0 (\varepsilon/\varepsilon_0)^{-\alpha},
\label{eq:n0}
\eeq
where $\alpha = 0$ for $\varepsilon < \varepsilon_0$ and $\alpha = 2$
for $\varepsilon > \varepsilon_0$ .  Here, $\varepsilon_0 = h
\nu_0^{ob}/\Gamma = 6.6 \times 10^{-14} \, \Gamma^{-1} $~erg. The
normalization constant, $n_0$ is found by normalizing to the total
bolometric luminosity $L_{\rm bol}^{ob} \equiv 10^{43} L_{\rm
  bol,43}^{ob} \, {\rm erg \, s^{-1}}$.  Note that this bolometric
luminosity does not correspond to the photon luminosity of the weak jets
themselves, which are necessarily weaker.  However, the survival of
UHE nuclei depends on the entire photon field.
Using the (comoving) energy density at the acceleration radius, $u =
L_{\rm bol}^{ob}/(4\pi r^2 \Gamma^2 \beta c) = \int n(\varepsilon)
\varepsilon d\varepsilon$, one finds
\beq
\ba{lcl}
n_0 & = & {L_{\rm bol}^{ob} \over 4 \pi \varepsilon_0^2 r^2 \Gamma^2 \beta
  c [1/2 + \log(\varepsilon_{\max}/\varepsilon_0)]} \nonumber \\
& \simeq & {6 \times
  10^{56} \over r^2 \Gamma^2 \beta} \, L_{\rm bol,43}^{ob}
{\nu_{0,13}^{ob}}^{-2} \; {\rm cm^{-3} \, erg^{-1}},
\label{eq:n0_b}
\ea
\eeq
where we approximated $\log(\varepsilon_{\max}/\varepsilon_0) \approx
10$.

For the photon spectrum given by equation \ref{eq:n0}, the inner
integral in equation \ref{eq:t_dis1} can be written as
\beq
\int_{\epsilon'/2 \gamma_A}^\infty {n(\varepsilon'') \over
  \varepsilon''^2} d\varepsilon'' = {n_0 /\varepsilon_0 \over (1+ \alpha)}
\left({\epsilon' \over 2 \gamma_A \varepsilon_0}\right)^{-(1+\alpha)}.  
\label{eq:integral1}
\eeq
(Note that the above result is accurate for $\epsilon'/2 \gamma_A >
\varepsilon_0$, while for  $\epsilon'/2 \gamma_A < \varepsilon_0$
there is a second term that can be neglected for large enough values
of $\gamma_A$; see further discussion below.) 

\begin{figure}
\includegraphics[width=\linewidth]{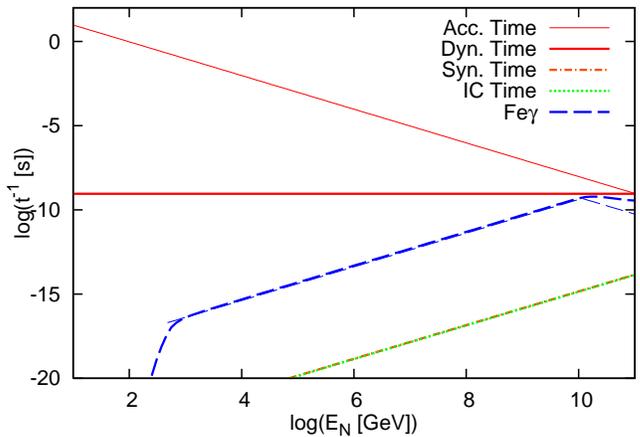}
\caption{ The acceleration time scale and various cooling time scales
  of iron for RQ AGNs. Used parameters are $L_{bol}^{ob}=10^{43} \;
  {\rm erg \, s^{-1}}$, $L_B \equiv \epsilon_B L =10^{43} \; {\rm erg
    \, s^{-1}}$, $\nu_0^{ob}=10^{13} \; {\rm Hz}$, photon spectral
  index $\alpha=0$ for $\nu < \nu_0$ and $\alpha = 2$ for $\nu >
  \nu_0$, acceleration radius $r = 10^{19} \; {\rm cm}$, jet velocity
  $\beta=0.3$, and $\eta= \beta^{-2} = 10$. Shown are the inverse of
  the relevant time scales.  For the parameters shown here, the
  acceleration time (thin red line) is much shorter than the dynamical
  time (thick red line) and the synchrotron (dash-dotted red line) and
  Compton (dashed green line) cooling times of iron nuclei at energies
  below $10^{20}$~eV. The thick, dash blue line shows the exact
  numerical results of the calculation of iron photodisintegration
  time scale. This calculation includes the nonresonant effect, and is
  similar to the ones presented in Ref. \cite{MINN08}. The thin blue line,
  which by large overlaps the thick blue line, shows the analytical
  result discussed here, using equation \ref{eq:t_dis_final}. Note
  that the parameter set chosen allows for survival of
  ultrahigh-energy iron nuclei up to $\sim 10^{20}$~eV.  }
\label{fig:RQ}
\end{figure}

In order to estimate the outer integral in equation \ref{eq:t_dis1},
we discriminate between photodisintegration and photomeson production
processes. In the photodisintegration process, the main contribution
to the outer integral in equation \ref{eq:t_dis1} is from photons in
the energy bandwidth of the giant dipole resonance, whose energy is
given by $\epsilon_{\rm GDR} = 42.65\times A^{-0.21}$~MeV for $A>4$
\cite{KT93}. Numerical fits to the experimental data gives the energy
bandwidth of the resonance $\Delta_{\rm GDR} = 8$~MeV, and the maximum
value of the cross section to be $\sigma_{0,A} = 1.45 A\times 10^{-27}
{\rm \, cm^2}$. Approximating the outer integral by the contribution
from the resonance, one finds
\beq
\ba{lcl}
t_{\rm dis}^{-1} & \simeq & {c \over 2 \gamma_A^2} \int_{\epsilon_{\rm th}^{A}}^\infty
d\epsilon' \epsilon' \sigma_{A}(\epsilon') {n_0 \over (1+\alpha)}
{\left( {\epsilon' \over 2 \gamma_A \varepsilon_0}
  \right)}^{-(1+\alpha)} \nonumber \\ 
& \simeq & 
{c n_0 \sigma_{0,A} \Delta_{\rm GDR} \over \gamma_A (1+\alpha) } 
{\left( 2 \gamma_A  \varepsilon_0 \over \epsilon_{\rm GDR} \right)}^{\alpha}.
\label{eq:t_dis_final}
\ea
\eeq
Note that this result is in good agreement with the more detailed
numerical calculation at $\gamma_A \lsim \epsilon_{\rm
  GDR}/2\varepsilon_0$ (see Ref. \cite {MINN08} and figure
\ref{fig:RQ}) \footnote{The contribution from the non-GDR-resonance
  (including quasi-deuteron and fragmentation processes) region
  becomes important at $\gamma_A \gtrsim \epsilon_{\rm
    GDR}/2\varepsilon_0$ for $\alpha \lesssim 1$.}.

The photodisintegration time can be compared to the dynamical time,
$r/\Gamma \beta c$ \footnote{More accurately, the photodisintegration
  time should be compared to the escape time scale, which is highly
  uncertain.  If the escape time scale is too long, UHECRs would lose
  their energy due to, e.g., the adiabatic cooling (whose time scale
  is typically $t_{\rm ad} \sim 3 t_{\rm dyn}$). Here we assume that
  $t_{\rm esc}$ is longer than but comparable to $t_{\rm dyn}$, hence
  UHECRs can escape from the source without significant adiabatic
  loss.}, to obtain a lower limit on the acceleration radius which
allows survival of the accelerated nuclei.  The result depends on the
target photon spectrum. For the assumed spectral index $\alpha \sim 2$
above $\nu_0^{ob}$, the dissipation time obtains its minimum value at
$\gamma_A \sim \epsilon_{\rm GDR}/2 \varepsilon_0$. For nuclei with
this value of the Lorentz factor, one finds the survival condition,
\beq
t_{\rm dyn} t_{\rm dis}^{-1} < 1 \leftrightarrow r \gsim 2.1 \times
10^{16} A^{1.21} \, L_{\rm bol,43}^{ob} \, {\nu_{0,13}^{ob}}^{-2} \,
\Gamma^{-4} \, \beta^{-2} \; {\rm cm} 
\label{eq:survive}.
\eeq 
For iron nuclei, this gives $r\gsim 3 \times 10^{18} \, L_{\rm
  bol,43}^{ob} {\nu_{0,13}^{ob}}^{-2} \, \Gamma^{-4} \, \beta^{-2} \;
{\rm cm}$.  Therefore, for 10 parsec-scale weak jets (assuming here
$\beta \sim 0.3$) of RQ AGNs, we can expect that heavy nuclei can not
only be accelerated but also survive without significant
photodisintegration.  We further note that for nuclei at higher
energies, numerical calculations (e.g., Ref. \cite{MINN08}) indicate
roughly comparable constraints on the acceleration radius. As at these
high energies ($\gsim \epsilon_{\rm GDR}/2\varepsilon_0 =
10^{19}$~eV) the simple analytical treatment may be insufficient given
the contribution from non-GDR resonance, we carried a numerical
calculation. We present in figure \ref{fig:RQ} a comparison of the
analytical approximation of the energy loss time to a more accurate
numerical calculation, as well as the cooling time scales due to
additional phenomena. While some deviation at the very high energies
exist, clearly the analytical approximation used here is very accurate
at lower energies, and does not deviate much from the exact numerical
solution above $10^{19}$~eV and below $10^{20}$~eV, for the parameters
values chosen.

The second energy loss channel for energetic nucleons or nuclei is 
the photomeson production. The main contribution to the outer integral
in equation \ref{eq:t_dis1} is from photons at energies $\epsilon_{\rm
  peak} \sim 0.3 \GeV$, where the cross section peaks at the
$\Delta$-resonance, whose width is $\Delta \epsilon \approx 0.2
\GeV$. For protons, the peak of the cross section at the
$\Delta$-resonance is $\sigma_{\rm peak} \approx 5 \times 10^{-28} \,
{\rm cm^2}$, and the inelasticity is $\kappa_{\rm peak} \sim 0.2$
\cite{Stecker68}.  For heavier nuclei, the cross section is roughly
proportional to $A$ \cite{Michalowski77}, while the inelasticity is
proportional to $A^{-1}$ (see Ref. \cite{Stecker68}). Repeating the same
calculation as for the photodisintegration process presented above,
one finds that the limitation on the acceleration radius arises for
particles at Lorentz factor $\gamma_A \sim \epsilon_{\rm peak}/2
\varepsilon_0$.  For these particles, we have $t_{\pi}^{-1} \simeq 
(2/(1+\alpha)) n_0
\varepsilon_0 c \sigma_{\rm peak} \kappa_{\rm peak} (\Delta
\epsilon/\epsilon_{\rm peak}) {(\gamma_A/ 0.5 \epsilon_{\rm peak}
  \varepsilon_0^{-1})}^{\alpha-1}$  \footnote{ In a similar way to
  photodisintegration process, at high energies $\gamma_A \gtrsim
  \epsilon_{\rm peak}/2\varepsilon_0$, the energy loss time differs
  from the simple estimate given here if $\alpha \lesssim 1$ due to
  multi pion production. However, the $\Delta$-resonance approximation
  used here is still valid at $\gamma_A \lesssim \epsilon_{\rm
    peak}/2\varepsilon_0$.}.  The requirement that $t_{\rm dyn}
t_{\pi}^{-1} < 1$ leads to a much looser constraint than that for
photodisintegration. Note that this condition does not depend on $A$
for a given $\gamma_A$. We thus conclude, that under conditions that
enable survival of heavy nuclei, photomeson production is inefficient
(see also Refs. \cite{AHST08,MINN08}).

The number density of particles at this distance, $n
\approx L/4\pi r^2 \Gamma^2 \beta m_p c^3 \approx 2 \times 10^{-2} \, L_{\rm bol,43}^{ob} \,
r_{18}^{-2} \, \Gamma^{-2} \, \beta^{-1} {\rm cm^{-3}}$, implies that spallation is not
important as an energy loss channel of the energetic particles.  The
rate for spallation process due to collision with other nuclei can
be estimated as
\beq
t_{\rm sp}^{-1} = \sigma_{\rm sp} n c,
\label{eq:14}
\eeq
where $\sigma_{\rm sp} = 5 \times 10^{-26} A^{2/3} {\, \rm cm^{2}}$ is the
cross section for spallation of a nucleus \cite{WRM08}. 
One can thus conclude that for acceleration at parsec scales, the
spallation loss time is thus much longer than the dynamical time.

\section{Energetics of RQ AGNs and efficiency of UHECRs acceleration}
\label{sec:3}

In the previous section, we have concluded that the conditions inside
the weak jets of RQ AGNs enable the acceleration and survival of 
UHE nuclei, if particle acceleration occurs at parsec scale radii. 
Such radii may be characteristic of internal or standing shocks
in a jet, or shocks produced by interaction of the jet and the ambient
medium.  In this section, we compare the energy budget in RQ AGNs with
the energy requirement for UHECRs acceleration.  

The observed flux of UHECRs on earth is $\mathcal{J} \approx 5 \times
10^{-16} {\rm \, m^{-2} \, s^{-1} \, sr^{-1}}$ (see Ref. \cite{BEH09}
and references therein).  The production rate of UHECRs within the GZK
horizon of $R_{\rm GZK} \approx 100$~Mpc \footnote{Note that similar
  distances are obtained for proton as well heavy nuclei composition
  of UHECRs. See, e.g., Ref. \cite{Der07}.} is therefore ${\dot
  \mathcal{N}} = 4 \pi J/R_{\rm GZK} \approx 1.7 \times 10^{36} {\rm
  \, Mpc^{-3} \, yr^{-1}}$. For energies above $10^{19.5}$~eV, this
implies energy production rate in energetic particles of ${\dot
  \mathcal{E}} \approx 8.5 \times 10^{43} \, {\rm erg \, Mpc^{-3} \,
  yr^{-1}}$. This value is basically in agreement with more detailed
estimates (see, e.g., Refs. \cite{BGG06, MT09}).

Following Refs. \cite {Cap99,Gal+06, MSPC09} (and references therein),
we consider a typical RQ AGN jet luminosity $L_{\rm jet} \sim
{10}^{42-43}~{\rm erg} {\rm\, s}^{-1}$. These values are based on
measurements in the optical band \cite{Cap99}; The typical radio
emission is much weaker, $\sim {10}^{38-39}~{\rm erg} {\rm\, s}^{-1}$,
due to both synchrotron self absorption and free-free absorption.
Although the value for $L_{\rm jet}$ is highly uncertain, we find this
luminosity to be equal to or somewhat fainter than the magnetic
luminosity required for acceleration of UHE nuclei (say, with $Z\geq
10$, see eq. \ref{eq:2}).  The number density of RQ AGNs in the local
universe is $\phi_{42} \sim {10}^{-3}~{\rm Mpc}^{-3}$ which is about
10 times larger than RL AGNs (see, e.g., Ref. \cite{UH01}).
Therefore, we expect that the energy input from RQ AGNs is $\sim 3
\times {10}^{46} \, {\rm erg \, Mpc^{-3} \, yr^{-1}}$.

The number density of RQ AGNs drops with their luminosity \cite{HMS05,
  TUV09}.  For example, the number density of AGNs with soft X-ray
luminosity, $L_{X} \simeq 10^{44} \, \rm{ erg \, s^{-1}}$, is
approximately $\phi_{44} \simeq 10^{-6} {\rm \, Mpc^{-3}}$. While at
present the jet luminosity is not easily deduced from the X-ray
luminosity measurements (e.g., absorption has a different effect on
measurements at the soft and hard X band, see Ref. \cite{Tue+08}), we
can still conclude that expected energy output from RQ AGNs jets
significantly exceeds the energy requirement for UHECRs sources.
Therefore, RQ AGNs jets are energetic enough to supply acceleration of
particles to ultrahigh energies.

The estimated number density of RQ AGNs, $\phi_{42} \sim
{10}^{-3}~{\rm Mpc}^{-3}$, implies that for isotropic distribution of
sources, the rate of energy production in UHE particles in a single
source should be roughly $\approx 3 \times 10^{39} {\rm \, erg
  \,s^{-1}}$ at ${10}^{19.5}~{\rm eV}$, in order for these objects to
be the sources of UHECRs (and assuming that no additional sources of
UHECRs exist). This value is less than a percent of the expected
typical jet luminosity of these objects.

This obtained value of $\sim {10}^{39.5}~{\rm \, erg \, s^{-1}}$,
while well within the energy limit of a typical RQ AGN, should be
considered as a lower limit. First, the estimated number density of RQ
AGNs does not take into account the possibility that in some of these
objects the jet luminosities may be insufficient (see further
discussion below). Second, the uncertainty that exists in the value of
the efficiency parameter $\eta$, as well as in the shock velocity
$\beta$ implies, via equation \ref{eq:2} that the required luminosity
for acceleration of iron nuclei to $10^{20}$~eV may be somewhat higher
than the value $10^{42}~{\rm \, erg \, s^{-1}}$ considered here, in
which case the number density of available sources may be lower.
Therefore, even though our scenario remains essentially viable, one
should keep in mind the high uncertainty that exists in the details of
the physical processes. Nonetheless, the values obtained here can give
a first approximation of the physical requirement from the source of
UHECRs. The fact that they are well within the known energy budget of
these objects leaves an ample margin to incorporate many of the
uncertainties in the values of the physical parameters.

The above estimate does not take into account acceleration of
particles to lower energies, which accompanies the production of
UHECRs. Therefore, the required cosmic-ray input per source depends on
the spectral index of the accelerated particles in the
source. Although very steep source spectra can lead to an `energy
crisis' (see, e.g., Refs. \cite{BGG06,MIN08}), RQ AGNs can be viable
sources if sufficiently flat spectra are achieved.  As both the jet
luminosity and number density are uncertain at present, the results
obtained here can perhaps serve as a guideline. Additional source of
uncertainty lies in the energy losses caused by shock waves formed if
a large number of charged particles escape from the source. While the
energy of the escaping particles is reduced, still the highest energy
particles, for which the escape time is the shortest, will be less
affected. The exact details of this effect depend on the exact
spectrum of the escaping particles, and are thus beyond the scope of
this manuscript. However, we point here that if only the highest
energy particles escape, this effect is not expected to play a
significant role. Then, the overall cosmic-ray energy spectrum may be
achieved as a superposition of contributions from many RQ AGNs with
different cosmic-ray luminosities and maximum energies.  Finally, our
model requires a significant fraction of heavy element composition in
the jet. As the jet material is ejected from the inner part of the
disk, existence of metals in the disk inevitable leads to their
existence in the jet, unless nuclei are disrupted in the jet base and
the accretion flow in the disk.

The constraints found in \S\ref{sec:2} above imply that UHECR
production is possible if the acceleration radius is of the order of
parsec, in order to avoid energy loss by photodisintegration. 
We note that the calculations carried in \S\ref{sec:2} considered the
bolometric luminosity, which is higher than the jet luminosity. As a
result, the constraints derived above are more restrictive
than the constraints that would have obtained by considering the
(uncertain) luminosity in the jet only.

Using the results derived in \S\ref{sec:2} we can obtain a lower limit
on the value of the magnetic field at the acceleration site if we
assume the acceleration radius $r$, and characteristic velocity,
$\beta$.  If UHECRs are composed of iron nuclei, assuming $\beta =
0.3$, equation \ref{eq:survive} requires acceleration radius of $r
\sim 10$~pc. At this radius, the Hillas condition (equation
\ref{eq:1}) implies $B \gsim 10^{-4} \, \eta \; \beta_{-1/2}^3 \,
{E_{20}^{ob}}\, L_{\rm bol,43}^{ob} \, {\nu_{0,13}^{ob}}^{-2} $~G.
For Carbon nuclei, the lightest nuclei that are consistent with the
restrictions on the luminosity obtained in equation \ref{eq:2}, the
minimum acceleration radius consistent with equation \ref{eq:survive}
should be at $5 \times 10^{18}$~cm. The magnetic field at this radius
should thus exceed $B \geq 3 \times 10^{-3} \, \eta \; \beta_{-1/2}^3
\, {E_{20}^{ob}}\, L_{\rm bol,43}^{ob} \, {\nu_{0,13}^{ob}}^{-2} $~G.
We note that for both scenarios, sub-Gauss values of the magnetic
field is consistent with the requirement that the cooling time is
shorter than the dynamical time (see equation \ref{eq:t_cool_dyn}), in
which case the spectrum of the escaping particles is similar to that
determined by the acceleration mechanism. We can thus constrain the
required value of the magnetic field at the acceleration site to be in
the range $10^{-3} - 1$~G.

This $\sim 1$~parsec acceleration radius derived here is consistent with the
observed scale of weak jets observed in nearby Seyfert galaxies
\cite{MWPG03,GBO04,UWTGM05,Gal+06,MARK07}.  At this radius, internal
interactions with the jets, or interaction of the jet with the
interstellar medium material are expected to produce shock waves, that
dissipate the jet kinetic energy. These shock waves are plausible
acceleration sites for energetic particles. The lower limits on the
values of the magnetic field derived above can be translated into lower
limits on the equipartition value of the magnetic field at the
acceleration site: using $B^2/8\pi = \epsilon_B u$ (see discussion
above equation \ref{eq:2}), one finds that for iron nuclei,
$\epsilon_B \geq 4 \times 10^{-3}$, while for carbon nuclei,
$\epsilon_B \geq 10^{-1}$. These values are both below equipartition,
and are consistent with values of the magnetic field expected to be
produced in shock waves.

\section{Observational Implications}
\label{sec:implications}

In the previous sections, it has been shown that RQ AGNs can be viable
sources of UHECRs, provided that the UHECRs composition is dominated by
heavier nuclei (e.g., iron) rather than protons. This claim is based
on two general considerations: (1) For a heavy nuclei composition, the
required magnetic luminosity is weak enough to enable RQ AGNs to
accelerate UHECRs; and (2) RQ AGNs are about ten times more numerous 
than RL AGNs, so that they can significantly contribute to the observed 
UHECR flux.

If the assumption presented here is correct, in addition to the
contribution of RQ AGNs to the observed UHECR flux, we expect a
contribution from RL AGNs. This is because the same acceleration
mechanism can be expected to work in the latter.  Although it is
difficult to estimate the relative contribution of these two classes
of objects, we expect that heavy nuclei (presumably, iron) from RQ
AGNs can mask the correlation that may be found between the arrival
direction of UHECRs and RL AGNs. In this respect, our scenario is
consistent with the recent ``disappointing model'' \cite{ABG09},
which does not predict a significant correlation of the most energetic
CR's with AGNs.

Nonetheless, the conditions within the jets of RL AGNs may enable
proton acceleration to ultra-high energies within these objects as
well.  Thus, if our scenario is correct, even if the correlation that
can be expected between the arrival direction of UHECRs and positions
of RL AGNs is reduced by the additional contribution from RQ AGNs,
some correlation between the arrival directions of UHE protons and the
positions of RL AGNs is still expected.  The strength of this
correlation depends on the efficiency of particle acceleration in RQ
and RL AGNs, and on the composition of particle acceleration in RL
AGNs: as our model allows, in principle, proton acceleration to UHE in
RL AGNs, the observed UHECRs flux may be composed of two distinctive
populations: heavy nuclei accelerated in RQ AGNs, and protons
accelerated in RL AGNs. A testable consequence of this idea is that
UHECRs whose arrival directions correlate with the position of RL AGNs
will show, on the average, lighter composition than UHECRs whose
arrival direction is not correlated with RL AGNs. Nonetheless, we
stress that this is highly uncertain, due to two main reasons: first,
RL AGNs may accelerate heavy nuclei to high energies as well; and
second, the efficiencies of particle acceleration and escape from
these sources are highly uncertain.

We can therefore conclude that our model does not rule out some
correlation between the arrival direction of UHECRs and positions of
RL AGNs. We note though, that the current observational status is
highly uncertain. While some correlation was reported
\cite{Abraham07, Abraham09}, its strength seems not to be fully
determined yet (the recent results reported by Ref. \cite{Abraham09}
indicate much weaker correlation than earlier reports
\cite{Abraham07}).  Therefore, future anisotropy search focusing on
UHE proton events should be very important in confirming this idea.

In our scenario, the detection of secondary gamma rays and neutrinos
from individual RQ AGNs whose luminosity is thought to be $L \sim
{10}^{42-43}~{\rm erg} {\rm s}^{-1}$, seems to be difficult. This is
due to the fact that, as we saw in \S\ref{sec:photodisintegration}, in
the weak jets in RQ AGNs photodisintegration is the dominant energy
loss channel, hence copious production of energetic $\pi$'s is
not expected. We leave the details of this calculation for future
work.  

On the other hand, the typical luminosity of jets of RL AGNs can be a
few orders of magnitude brighter than RQ AGNs. Therefore, RL AGNs are
more favorable targets for the purpose of detecting energetic
secondaries. In particular, secondaries originating from very powerful
AGNs like FR II galaxies (see, e.g., Ref. \cite{AD09} and references
therein) or very nearby RL AGNs such as Cen A (e.g.,
Ref. \cite{DRFA08} and references therein) might be detected by Fermi
or IACTs.  Clusters of galaxies hosting powerful AGNs might also be
viable candidate for secondary detection surveys (see, e.g.,
Refs. \cite{MIN08,Kot+09} and references therein).

\section{Summary and Discussions}
\label{sec:summary}

In this paper we have considered radio quiet AGN's as possible sources
of UHECRs. So far, jets in these objects were not considered as
sources of the highest energy cosmic rays around $10^{20}$~eV, since
they are not luminous enough to support acceleration of protons to
these energies. However, the recent results of the AUGER collaboration
shows indications that at the highest energies the composition of
cosmic rays may be dominated by heavy nuclei \cite{Unger07,Belido09}.
The assumption that UHECRs are heavy nuclei eases the constraint on
the source luminosity (see eq. \ref{eq:2}). Thus it allows, in
principle, radio-quiet AGNs to be the sources of UHECRs.

We have calculated in \S\ref{sec:2} the constraints on the
acceleration site which enable the acceleration of heavy nuclei to
high energies. The most restrictive constraint arises from the
photo-disintegration process. We showed in
\S\ref{sec:photodisintegration}, that for typical nearby AGN's with
bolometric luminosity $L_{bol}^{ob} \approx 10^{43} \; {\rm erg \,
  s^{-1}}$, energetic nucleus can survive photo-disintegration if the
acceleration takes place on a parsec scale (see
eqs. \ref{eq:survive}). Interestingly, this is the same scale at which
weak jets were seen in these objects
\cite{MWPG03,GBO04,UWTGM05,MARK07}. This fact further supports our
idea, since interaction of jets with the surrounding material
inevitably leads to creation of shock waves, which are the most
plausible acceleration site of particles to high energies.

The question of the acceleration of particles to ultrahigh energies may
depend on the shock velocity. In the Bohm approximation $\eta
\propto \beta^{-2}$, some assumptions about the shock velocities in
the jets are needed. For example, a shock velocity of $\beta \sim 0.1$
requires magnetic luminosity of $\epsilon_B L \gsim 10^{43.5} \; {\rm
  erg \, s^{-1}}$ in order to enable acceleration of iron nuclei to
the maximum observed energy (see \S\ref{sec:2}, equation \ref{eq:2}),
which may be too low. However, somewhat higher values, $\beta \sim
0.3$ may be sufficient. Since currently there is a high uncertainty in
the determination of the shock velocities in RQ AGNs, it is not possible to 
put further constraints. We have also showed in \S\ref{sec:3} that the
minimum inferred values of the magnetic field required to confine the
accelerated particles at parsec scale jets is sub-Gauss. This value
was shown to be smaller than the equipartition value, which is also
consistent with models of particle acceleration in shock waves that
may be produced on this scale.

We further showed in \S\ref{sec:3}, that the required luminosity in
UHECRs could be less than a percent of the total bolometric luminosity
of more abundant nearby AGNs. If the spectrum of produced cosmic rays
is a power law with index close to $d \log N/d\log E \approx -2$, as
suggested by models of particle acceleration in non-relativistic shock
waves \cite{BE87}, then the total luminosity in energetic particles
(at all energies) could still be much smaller than the total
bolometric luminosity. Even for higher power law index, $p=2.3$, we
find that the total energy requirement for acceleration of cosmic rays
above $\GeV$ is $\approx (10^{10})^{0.3} = 10^3$ times higher than the
energetic requirement from UHECRs alone, which is still (marginally)
consistent with the total energy budget in RQ AGNs (see discussion in
\S\ref{sec:3}).  Assuming that the main radiative source in these
objects is the accretion disk, this implies a high efficiency, of the
order of tens of percents, in the conversion of accretion energy to
acceleration of particles. Since large uncertainties exist in the
efficiencies of both the energy conversion in the jet and the
acceleration of particles, we can only conclude that our model is
consistent with the energetic requirements, provided a high efficiency
is achieved in both these processes.

Spectral synthesis as well as chemical enrichment models predict that
AGN's are metal rich. The metallicities in the broad emission line
region is typically $\sim 1$ to $\gsim 10$ times the solar
metallicity \cite{HF93, Ham02}, and grows with the luminosity
\cite{Shemmer04}. Thus, abundant existence of heavy nuclei in AGN's
disks is expected. As the plasma jet is composed of material from the
disk, abundant population of heavy nuclei in the jet is
expected. Thus, AGN's may be one of the natural sources of high energy
heavy nuclei.


If indeed the acceleration takes place on a parsec scale, then the
main energy loss channel is photodisintegration, rather than
photomeson production. As discussed in \S\ref{sec:implications}, this
fact implies that we do not expect abundant production of energetic
neutrinos, that may result from the decay of energetic pions. On the
contrary: the results in equations \ref{eq:survive} indicate that the
relative contribution of photomeson production may be no more than few
percents of photodisintegration as an energy loss channel of
UHECRs. Thus, we do not expect a copious production of neutrinos under
the conditions which enable the acceleration and survival of heavy
nuclei, as considered in this paper.

After they escape from their sources, UHECR nuclei can be subject to
photodisintegration in the intergalactic space. UHE nuclei with energy
$\gsim 10^{19}$~eV mainly interact with the cosmic infrared
background (CIB) photons \cite{SS99,Allard05, Allard08}. UHE iron
nuclei with energy $\lsim 10^{20}$~eV have a mean free path of
$\gsim 500$~Mpc, while the mean free path of oxygen nuclei at a
similar energy is only $\sim 30$~Mpc \cite{HST07}. Both increase
rapidly as the energy of the nuclei decreases. Thus, the GZK horizon
of UHE nuclei is comparable to, or even somewhat larger than the GZK
horizon of energetic protons (see also Ref. \cite{Der07}).

To summarize, we have pointed out here that the assumption that UHECRs
are composed of heavy nuclei, as is suggested by recent AUGER results
\cite{Unger07, Belido09}, enables radio quiet AGN's to be their
sources. This picture is consistent with what is currently known about
the existence of parsec scale jets seen in some of these objects, as
well as the energy constraints. Moreover, this picture is supported by
the recent analysis carried by Ref. \cite{George+08}, in which a
strong correlation between the arrival directions of UHECRs and the sky
coordinates of AGN's detected by the {\it Swift}-BAT within the GZK
cutoff (which are largely radio-quiet) was found.  As more data are
collected by the AUGER collaboration, a clearer picture of the
composition of UHECR's will become available, allowing further
constraints on possible sources.

\acknowledgments
We would like to thank the referee for his/hers useful comments.
AP wishes to thank Tim Heckman, Julian Krolik, Ohad Shemmer and Lukasz
Stawarz for useful discussions.  This work is partially supported by
the NORDITA program on Physics of relativistic flows, during which
part of this research was carried. AP is supported by the Riccardo
Giacconi fellowship award of the Space Telescope Science Institute.
KM acknowledges support by a Grant-in-Aid from JSPS and
by a Grant-in-Aid for the Global COE Program ``The
Next Generation of Physics, Spun from Universality and
Emergence'' from MEXT. This work was partially supported by NSF
PHY-0757155 and NASA NNX 08AL40G.


\begin{thebibliography}{srt}

\bibitem {Der07}  
 Dermer, C.D., Proc. 30th Intl. Cosmic Ray Conference (Merida,
 Mexico), in press (arXiv:0711.2804) (2007) 

\bibitem {BEH09}
 Bl\"umer, J., Engel, R., \& H\"orandel, J.R., Progress in
 Particle and Nuclear Physics {\bf 63}, 293 (2009)

\bibitem {Greisen66}
 Greisen, K., Phys. Rev. Lett. {\bf 16}, 748 (1966)

\bibitem{ZK66}
 Zatsepin, G.T., \& Kuzmin, V.A., JETP {\bf 4}, 78 (1966)

\bibitem {Tak98}
 Takeda, M. \etal, Phys. Rev. Lett. {\bf 81}, 1163 (1998)

\bibitem {Abbasi08}
 Abbasi, R.U. \etal, Phys. Rev. Lett. {\bf 100}, 101101 (2008)

\bibitem {Abraham08b}
 Abraham,  J. \etal  (The Pierre Auger Collaboration), Phys. Rev. Lett.
 {\bf 101}, 061101 (2008b)

\bibitem {Hillas84}
 Hillas, A.M., ARA\&A {\bf 22}, 425 (1984)

\bibitem {MU95}
 Milgrom, M., \& Usov, V., Astrophys. J. {\bf 449}, L37 (1995)

\bibitem{W95}
 Waxman, E., Phys. Rev. Lett. {\bf 75}, 386 (1995)

\bibitem {Vietri95}
 Vietri, M., Astrophys. J. {\bf 453}, 883 (1995)

\bibitem {Der02}  
 Dermer, C.D., Astrophys. J. {\bf 574}, 65 (2002) 

\bibitem{W04}
 Waxman, E., Astrophys. J. {\bf 606}, 988 (2004)

 \bibitem {MINN08}
 Murase,K., Ioka, K., Nagataki, S., \& Nakamura, T.,
 Phys. Rev. D {\bf 78}, 023005 (2008a)

\bibitem {MINN06}
 Murase,K., Ioka, K., Nagataki, S., \& Nakamura, T.,
 Astrophys. J. {\bf 651}, L5 (2006)
 
\bibitem {WRMD07}
Wang, X.,-Y., Razzaque, S., M\'esz\'aros, P. \& Dai, Z.-G., 
Phys. Rev. D {\bf 76}, 083009 (2007)

\bibitem {BS87}
 Biermann, P.L., \& Strittmatter, P.A., Astrophys. J. {\bf 322}, 643 (1987)

\bibitem {RB93}
 Rachen, J.P., \& Biermann, P.L., Astron. Astrophys. {\bf 272}, 161 (1993)

\bibitem {NMA95}
 Norman, C.A., Melrose, D.B., \& Achterberg, A., Astrophys. J. {\bf 454}, 60 (1995)

\bibitem {Aha02}
Aharonian, F.A., Mon. Not. R. Astron. Soc. {\bf 332}, 21 (2002)

\bibitem {BGG06}
 Berezinsky, V., Gazizov, A., \& Grigorieva, S., Phys. Rev. D {\bf 74}, 043005 (2006)

\bibitem {Berezhko08}
 Berezhko, E.G., Astrophys. J. {\bf 684}, L69 (2008)
 
\bibitem {DRFA08}
 Dermer, C.D., Razzaque, S., Finke, J.D., \& Atoyan, A., New
 Journal of physics {\bf 11}, 065016 (2008)
 
\bibitem {PS92}
 Protheroe, R.J., \& Szabo, A.P., Phys. Rev. Lett. {\bf 69}, 2885 (1992)

\bibitem {BG99}
Boldt, E., \& Ghosh, P. Mon. Not. R. Astron. Soc., {\bf 307}, 491 (1999)

\bibitem {Arons03}
 Arons, J., Astrophys. J. {\bf 589}, 871 (2003)
 
 \bibitem{MMZ09}
 Murase, K., M\'esz\'aros, P., \& Zhang, B., 
Phys. Rev. D {\bf 79}, 103001 (2009)
    
\bibitem {KRJ96}
 Kang, H., Ryu, D., \& Jones, T.W., Astrophys. J. {\bf 456}, 422 (1996)

\bibitem {ISMA07}
 Inoue, S., Sigl, G., Miniati, F., \& Armengaud, E., preprint (astro-ph/0701167) (2009)
  
\bibitem{VCV06}
 V\'eron-Cetty, M.-P., \& V\'eron, P., Astron. Astrophys. {\bf 455}, 773 (2006)

\bibitem {Abraham07}
 Abraham,  J. \etal  (The Pierre Auger Collaboration), Science {\bf 318}, 938 (2007)

\bibitem {Abraham08a}
 Abraham,  J. \etal  (The Pierre Auger Collaboration), 
Astropart. Phys. {\bf 29}, 188 (2008a)

\bibitem {KS08}
 Kashti, T., \& Waxman, E., J. Cosmol. Astropart. Phys. {\bf 5}, 6 (2008)

\bibitem {Tak08}
 Takami, H., Nishimichi, T., Yahata, K., \& Sato, K., 
J. Cosmol. Astropart. Phys. {\bf 6}, 31 (2009)

\bibitem {Ghis08}
 Ghisellini, G. \etal, Mon. Not. R. Astron. Soc. {\bf 390}, L88 (2008)

 \bibitem {Abbasi08b}
 Abbasi, R.U. \etal, Astropart. Phys. {\bf 30}, 175 (2008)

\bibitem {Abraham09}
 Abraham,  J. \etal  (The Pierre Auger Collaboration), in
 proc. of ICRC31 (arXiv:0906.2347) (2009)

\bibitem {Unger07}
 Unger, M. \etal (Pierre Auger Collaboration), Astronomische Nachrichten {\bf 328}, 614
(arXiv:0706.1495 (2007))

\bibitem {Belido09}
 Bellido, J.A. \etal (Pierre Auger Collaboration), in Proc. of the XXth Rencontres de Blois 2008 "Challenges in Particle Astrophysics" (arXiv:0901.3389) (2009)

\bibitem {Kro94}
 Kronberg P.-P., Rep. Prog. Phys. {\bf 57}, 325 (1994)

\bibitem {RKCD08}
Ryu, D., Kang, H., Cho, J., \& Das, S., Science {\bf 320}, 909 (2008)

\bibitem {DKRC08}
Das, S., Kang, H., Ryu, D., \& Cho, J., Astrophys. J. {\bf 682}, 29 (2008) 

\bibitem {MSPC09}
Moskalenko, I.~V., Stawarz, L., Porter, T.~A., \& Cheung, C.~C.,
 Astrophys. J. {\bf 693}, 1261 (2009)

\bibitem {TS09}
 Takami, H., \& Sato, K., Astropar. Phys. {\bf 30}, 306 (2009)

\bibitem{ZFG09}
 Zaw, I., Farrar, G.R., \& Greene, J.E., Astrophys. J. {\bf 696}, 1218 (2009)

\bibitem {FG09}
 Farrar, G.R., \& Gruzinov, A., Astrophys. J. {\bf 693}, 329 (2009)

\bibitem {Sig09}
Sigl, G., New Journal of Physics {\bf 11}, 065014 (2009)

 \bibitem{MT09}
 Murase,K., \& Takami, H., Astrophys. J. {\bf 690}, L14 (2009)

\bibitem{WL08}
 Waxman, E., \& Loeb, A., J. Cosmol. Astropart. Phys. {\bf 8}, 26 (2009)

\bibitem {Allard05}
 Allard, D., Parizot, E., Olinto, A.V., Kahn, E., \& Goriely, S., 
Astron. Astrophys. {\bf 443}, L29 (2005)

\bibitem {Allard08}
 Allard, D., Busca, N.G., Decerprit, G., Olinto, A.V., \&  Parizot,
 E., J. Cosmol. Astropart. Phys. {\bf 10}, 33 (2008)

\bibitem {ABG09}
Alosio, R., Berezinsky, V.S., Gazizov, A., arXiv0907.5194 (2009)

\bibitem {PC09}
 Pe'er, A., \& Casella, P., Astrophys. J. {\bf 699}, 1919 (2009)

\bibitem {MWPG03}
 Mundell, C.G., Wrobel, J.M., Pedlar, A., \& Gallimore, J.F.,
 Astrophys. J. {\bf 583}, 192 (2003)

\bibitem {GBO04}
 Gallimore, J.F., Baum, S.A., \& O'Dea, C.P., Astrophys. J. {\bf 613}, 794 (2004)

\bibitem{UWTGM05}
 Ulvestad, J.S., Wong, D.S., Taylor, G.B., Gallimore, J.F., \&
 Mundell, C.G., Astrophys. J. {\bf 130}, 936 (2005)

\bibitem {Gal+06}
 Gallimore, J.F. \etal, Astrophys. J. {\bf 132}, 546 (2006)

\bibitem {MARK07}
 Middelberg, E., Agudo, I., Roy, A.L., \& Krichbaum, T.P.,
 Mon. Not. R. Astron. Soc. {\bf 377}, 731 (2007)

\bibitem {Ho08}
 Ho, L.C., Ann. Rev. Astro. Astrophysics {\bf 46}, 475 (2008)
 
\bibitem {LO07}
Lyutikov, M., \& Ouyed, R., Astroparticle Physics {\bf 27}, 473 (2007)

\bibitem {AM04}
 Alvarez-Mu\~niz, J., \& M\'esz\'aros, P., Phys. Rev. D. {\bf 70}, 123001 (2004)

\bibitem {Drury83}
 Drury, L. Oc., Reports on Progress in Physics {\bf 46}, 973 (1983)

\bibitem {BE87}
 Blandford, R.D., \& Eichler, D., Phys. Rep. {\bf 154}, 1 (1987)
 
\bibitem {Abdo+09}
 Abdo, A.A. \etal (the Fermi Collaboration), Astrophys. J. {\bf 700}, 597 (2009)

\bibitem{WRM08}
 Wang, X.,-Y., Razzaque, S., \& M\'esz\'aros, P., Astrophys. J. {\bf 677}, 432 (2008) 

\bibitem {BRS90}
 Begelman, M.C., Rudak, B., \& Sikora, M., Astrophys. J. {\bf 362}, 38 (1990)

\bibitem {CZS92}
 Chodorowski, M.J., Zdziaski, A.A., \& Sikora, M., Astrophys. J. {\bf 400}, 181 (1992)

\bibitem {MPR01}
 Mannheim, K., Protheroe, R.J., \& Rachen, J.P., Phys. Rev. D {\bf 63}, 023003 (2001)

\bibitem {AHST08}
 Anchordoqui, L.A., Hooper, D., Sarkar, S., \& Taylor, A.M., Astropar. Phys. {\bf 29}, 1 (2008) 

\bibitem {PSB76}
 Puget, J.L., Stecker, F.W., \& Bredekamp, J.H., Astrophys. J. {\bf 205}, 638 (1976)

\bibitem{Stecker68}
 Stecker, F.W., Phys. Rev. Lett. {\bf 21}, 1016 (1968)

\bibitem{Stecker69}
 Stecker, F.W., Phys. Rev. {\bf 180}, 1264 (1969)

\bibitem {Elvis94}
 Elvis, M. \etal, Astrophys. J. Supl. {\bf 95}, 1 (1994)

\bibitem {RR05}
 Rowan-Robinson, M. \etal, Astrophys. J. {\bf 129}, 1183 (2005)

\bibitem {Richards06}
 Richards, G.T. \etal, Astrophys. J. Supl. {\bf 166}, 470 (2006)

\bibitem {KT93}
 Karakula, S., \& Tkaczyk, W., Astropar. Phys. {\bf 1}, 229 (1993)

\bibitem {Michalowski77}
 Michalowski, S., Andrews, D., Eickmeyer, J., Gentile, T., Mistry, N.,
 Talman, R., \& Ueno, K., Phys. Rev. Lett. {\bf 39}, 737 (1977)

\bibitem {Cap99}
 Capetti, A., Axon, D. J., Macchetto, F. D., Marconi, A., \& Winge,
 C., Astrophys. J. {\bf 516}, 187 (1999)

\bibitem{UH01}
Ulvestad, J.S., \& Ho, L.S., Astrophys. J. {\bf 558}, 561 (2001)

\bibitem {HMS05}
 Hasinger, G., Miyaji, T., \&  Schmidt, M., Astron. Astrophys. {\bf 441}, 417 (2005)

\bibitem {TUV09}
 Trister, E., Urry, C.M., \& Virani, S., Astrophys. J. {\bf 696}, 110 (2009)

\bibitem {Tue+08}
Tueller, J. \etal, Astrophys. J. {\bf 681}, 113 (2008)

 \bibitem {MIN08}
 Murase, K., Inoue, S., \& Nagataki, S., Astrophys. J. {\bf 689}, L105 (2008b)
 
 \bibitem {AD09}
 Atoyan, A, \& Dermer, C.D., Astrophys. J. {\bf 687}, L75 (2009)

\bibitem {Kot+09}
 Kotera, K. \etal, arXiv:0907.2433 (2009)
 
\bibitem {HF93}
 Hamann, F., \& Ferland, G., Astrophys. J. {\bf 418}, 11 (1993)

\bibitem {Ham02}
 Hamann, F., Korista, K.T., Ferland, G.J., Warner, C., \& Baldwin, J., Astrophys. J. {\bf 564}, 592 (2002)

\bibitem {Shemmer04}
 Shemmer, O., Netzer, H., Maiolino, R., Oliva, O., Croom, S., Corbett,
 E., \& Fabrizio, L., Astrophys. J. {\bf 614}, 547 (2004)

\bibitem {SS99}
 Stecker, F.W., \& Salamon, M.H., Astrophys. J. {\bf 512}, 521 (1999)

\bibitem {HST07}
 Hooper, D., Sarkar, S., \& Taylor, A.M., Astropar. Phys. {\bf 27},
 199 (2007)

\bibitem {George+08}
George, M.R., Fabian, A.C., Baumgartner, W.H., Mushotzky, R.F., \& Tueller, J.,
 Mon. Not. R. Astron. Soc. {\bf 388}, L59 (2008)








\end{thebibliography}
\end{document}